\documentclass[11pt]{cernrep}
\usepackage{epsfig}
\usepackage{bm}

\begin{document}

\title{3-flavour lattice QCD at finite density and temperature:\\
QCD at finite isospin density revisited%
\thanks{$\;$ Talk presented by D.~K.~Sinclair.}}

\author{J.~B.~Kogut$^{1}$\thanks{$\;$ Supported in part by NSF grant NSF 
PHY03-04252.}
and D.~K.~Sinclair$^{2}$\thanks{$\:$This work was supported by the U.S.
Department of Energy, Division of High Energy Physics, Contract 
\mbox{W-31-109-ENG-38}.}}

\institute{$^1$Department of Energy, Division of High Energy Physics, 
Washington, DC 20585, USA\\
and\\
Dept. of Physics -- TQHN, Univ. of Maryland, 82 Regents Dr., College
Park, MD 20742, USA\\
$^2$ HEP Division, Argonne National Laboratory, 9700 South Cass Avenue, Argonne,
IL 60439, USA}

\maketitle

\begin{abstract}
We simulate 3-flavour lattice QCD at finite temperature and isospin chemical
potential $\mu_I$. In particular we study the nature of the finite temperature
transition for quark masses close to the critical mass at which this transition
for zero chemical potentials changes from a first order transition to a
crossover. We find that the Binder cumulants, used to determine the position of
this transition, have very strong $dt$ dependence. This leads us to an estimate
of the critical mass which is about $20\%$ below previous estimates. In
addition, when this $dt$ dependence is taken into account, we find that the
Binder cumulants show very little dependence on $\mu_I$. From this we conclude 
that we do not as yet see any evidence for the expected critical endpoint. We
have argued previously that the position and nature of the finite temperature 
transition for small $\mu_I$ should be the same as that for small quark-number
chemical potential $\mu$.
\end{abstract}  

\section{Introduction}

QCD at finite baryon-number density and temperature was present in the early
universe and is produced in relativistic heavy-ion colliders. QCD at finite
baryon-number density and zero temperature describes nuclear matter, such as
is found in neutron stars.

Because QCD at a finite chemical potential $\mu$ for quark number has a complex
determinant, standard simulation methods, based on importance sampling, fail. At
small $\mu$, in the neighbourhood of the finite temperature transition, various
approaches have been devised to circumvent these problems. These methods
include reweighting \cite{Fodor:2001au},
analytic continuation 
\cite{deForcrand:2002ci,D'Elia:2002gd,Azcoiti:2004ri}
and series expansions \cite{Allton:2002zi,Gavai:2003mf}. 
We adopt a
simpler approach and simulate with at a finite isospin chemical potential
$\mu_I$ \cite{Kogut:2004zg}. 
This corresponds to simulating with finite quark-number chemical
potential $\mu=\mu_I/2$ using only the magnitude of the fermion determinant and
ignoring its phase. For $\mu$ sufficiently small, this phase is well enough 
behaved on the sizes of lattice needed for the simulations, that one might
expect that the position and nature of the transitions at finite $\mu$ and
those at finite $\mu_I$ should be the same.

For 3 flavours of quarks, at zero chemical potentials, there is a critical
mass $m_c$ such that the finite temperature transition is first order for
$m < m_c$, second order in the 3-d Ising model universality class at $m=m_c$
and becomes a crossover for $m > m_c$ \cite{Karsch:2001nf} (see also
\cite{Christ:2003jk,deForcrand:2002ci}). 
Here it was expected that, as $|\mu|$ or
$|mu_I|$ was increased, $m_c$ would also increase, providing a critical
endpoint at some small chemical potential.

We are performing simulations with $m$ close to $m_c$ on $8^3 \times 4$ and
$12^3 \times 4$ lattices. What we find is that the fourth order Binder
cumulants, which are used to determine the nature of the transition, are very
sensitive to the updating increment $dt$ used in the hybrid molecular-dynamics
(R algorithm) simulations \cite{Kogut:2005qg} 
When this dependence is taken into account, 
$m_c \lappeq 0.027$ compared with previous estimates of $m_c \approx 0.033$.
In addition we have determined that the Binder cumulants have very little
$\mu_I$ dependence. Thus, as yet, we have obtained no evidence for a critical 
endpoint. Both these observations are in agreement with the new results of 
de Forcrand and Philipsen \cite{Philipsen}.

In section~2 we describe our simulations and presents our results. Discussions 
and conclusions are presented in section~3.

\section{Simulations and Results}

\begin{figure}[htb]
\epsfxsize=4in
\centerline{\epsffile{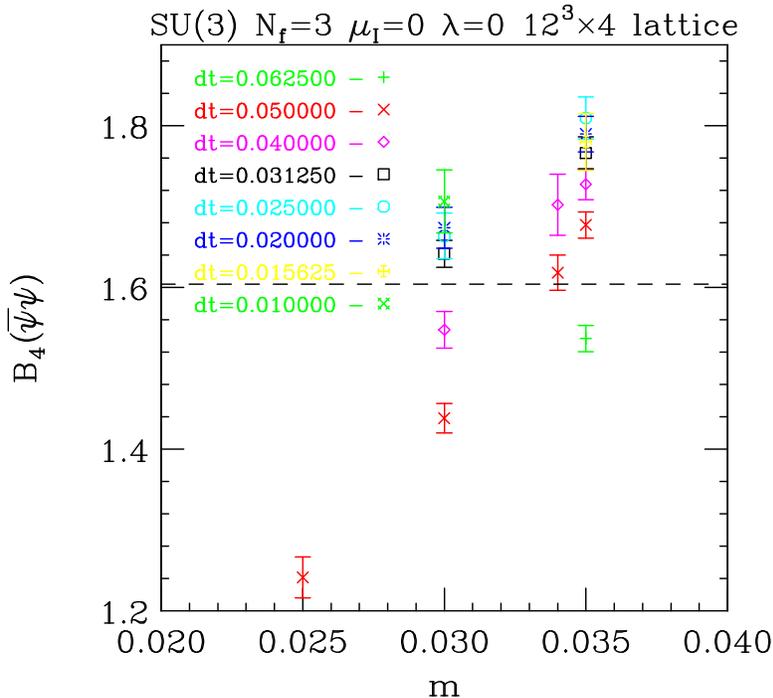}}
\caption{Mass dependence of $B_4(\bar{\psi}\psi)$ for various values of the
updating increment $dt$.}
\label{fig:b4m}
\end{figure}
\begin{figure}[htb]
\epsfxsize=4in
\centerline{\epsffile{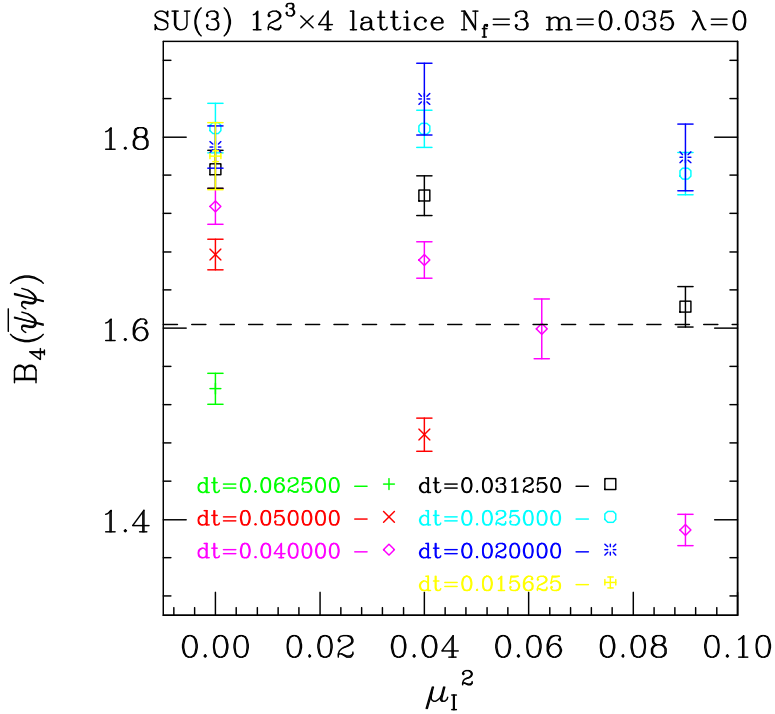}}
\caption{$\mu_I$ dependence of $B_4(\bar{\psi}\psi)$ for various values of the
updating increment $dt$ at $m=0.035$.}
\label{fig:b4mu}
\end{figure} 
\begin{figure}[htb]
\epsfxsize=4in
\centerline{\epsffile{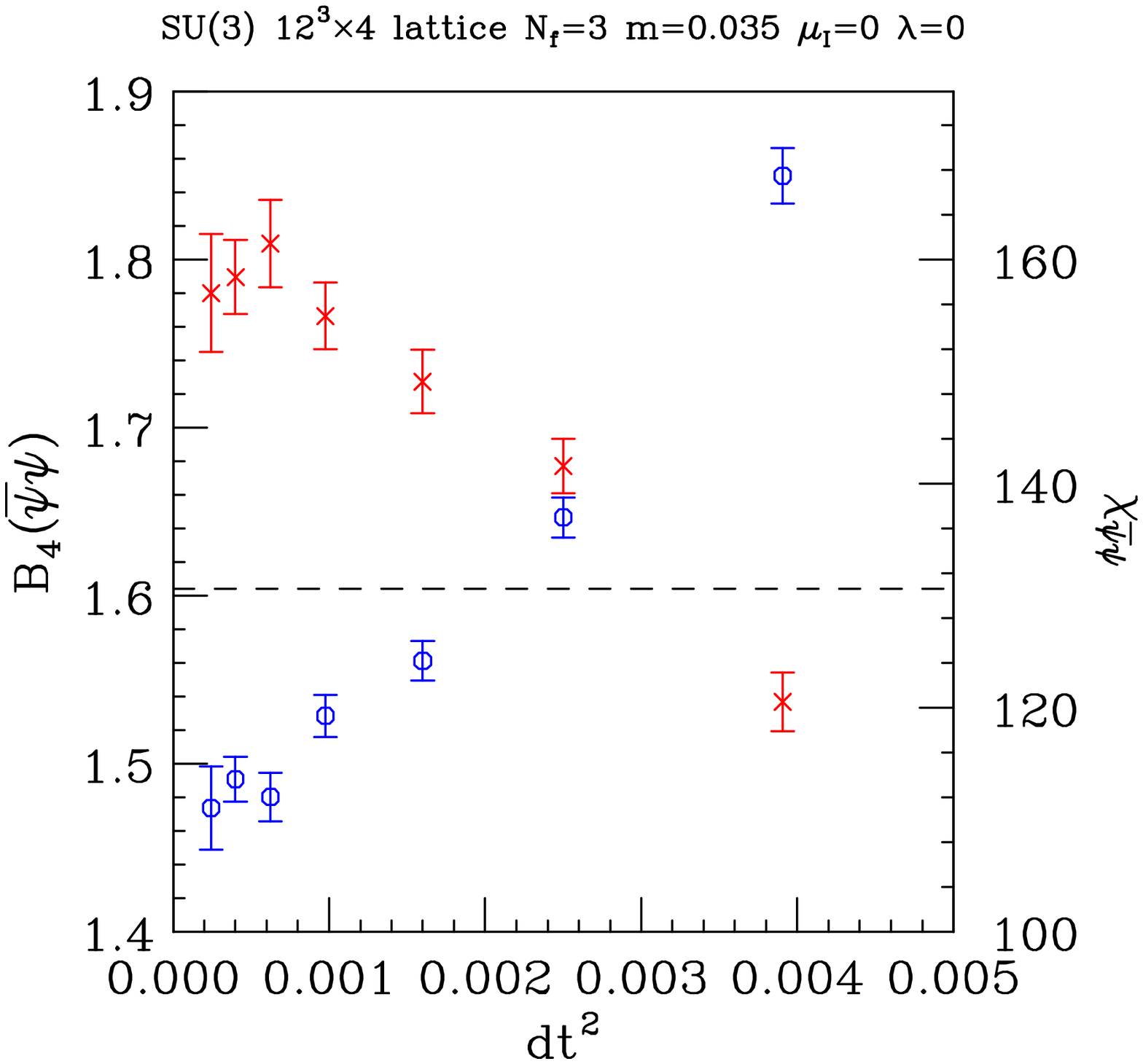}}
\caption{$dt$ dependence of $B_4(\bar{\psi}\psi)$ and $\chi_{\bar{\psi}\psi}$
at $m=0.035$ and $\mu_I=0$. The crosses are the Binder cumulants the circles
are the susceptibilities.}
\label{fig:m035mu0}
\end{figure}

The staggered quark action for lattice QCD at finite $\mu_I$ is
\begin{equation}
S_f={\sum_{sites}}\left\{\bar{\chi}[D\!\!\!\!/(\frac{1}{2}\tau_3\mu_I)+m]\chi
    + i\lambda\epsilon\bar{\chi}\tau_2\chi\right\}.
\end{equation}
which yields a real positive fermion determinant. We simulate this theory 
with 3 flavours of light dynamical quarks on $8^3 \times 4$ and $12^3 \times 4$
lattices using hybrid molecular-dynamics (R algorithm) methods, in the
viscinity of the finite temperature transition. Since we are interested in
small $\mu_I$ ($\mu_I < m_\pi$) we set $\lambda=0$.

We run at several quark masses in the range $0.025 \le m \le 0.04$, that is,
close to the critical mass. The position of the transition is determined by
the minimum of the fourth-order Binder cumulant \cite{binder}
for the chiral condensate,
defined by
\begin{equation}
B_4(\bar{\psi}\psi) =
{\langle(\overline{\bar{\psi}\psi}
-\langle\overline{\bar{\psi}\psi}\rangle)^4\rangle \over
 \langle(\overline{\bar{\psi}\psi}
-\langle\overline{\bar{\psi}\psi}\rangle)^2\rangle^2},
\end{equation}
where the overline indicates an average over the lattice. 5 noisy estimators of
$\overline{\bar{\psi}\psi}$ are made at the end of each trajectory, which yields
an unbiased estimator for $B_4(\bar{\psi}\psi)$. We have checked that
this method for determining the position of the transition produces an estimate
which is consistent with the position of the maximum of the corresponding
susceptibility. This measurement is obtained by simulating at typically 4
values of $\beta=6/g^2$ in the neighbourhood of the transition for a given
mass, $\mu_I$ (and $dt$), and continuing to the minimum using
Ferrenberg-Swendsen reweighting \cite{Ferrenberg:yz}. 
These simulations are usually 160,000 
length-$1$ trajectories long for each $(m,\beta,\mu_I)$.

Because $B_4$ depends strongly on the updating increment $dt$, we have
performed simulations for several different $dt$s in the range $0.01 \le dt
\le 0.0625$ for each $(m,\mu_I)$. Figure~\ref{fig:b4m} shows the Binder
cumulants as functions of mass for the various $dt$ values used, for our $12^3
\times 4$ runs. As we can see, the $dt$ dependence is considerable. At the
critical mass, $B_4$ should take the Ising value ($1.604(1)$). If we estimate
this from our $dt=0.05$ simulations we would conclude that $m_c \approx
0.0335$. If on the other hand we use our $dt=0.02$ runs we predict $m_c
\approx 0.027$. Thus we conclude that $m_c \lappeq 0.027$.

Figure~\ref{fig:b4mu} shows the $\mu_I$ dependence of $B_4$ for $m=0.035$ 
over the range $0 \le \mu_I \le 0.3$ ($m_\pi \sim 0.4$), also on a 
$12^3 \times 4$ lattice. What we see is that, for larger $dt$ values, $B_4$
appears to decrease with increasing $\mu_I$, which would lead us to the
erroneous conclusion that there was a critical endpoint where $B_4$ passes
through the Ising value for some $\mu_I < 0.2$. As $dt \rightarrow 0$, this
falloff becomes less pronounced, and there is even the possibility that
$B_4$ increases with increasing $\mu_I$. Thus we find no evidence for a 
critical endpoint.

In figure~\ref{fig:m035mu0}, we plot the values of $B_4$ at $m=0.035$ and 
$\mu_I=0$ against $dt^2$. From this it is clear that, at the smallest $dt$s,
$B_4$ is close to its $dt=0$ value. As we go to larger $dt$, $B_4$ decreases
making the transition appear more abrupt, i.e. less like a crossover and more
like a first-order transition. Also included in this figure are the chiral
susceptibilities, showing their strong $dt$ dependence. These increase with
increasing $dt$, again making the transition appear more abrupt.

The transition $\beta$, $\beta_c$ and hence the transition temperature 
decrease slowly as $\mu_I$ is increased. Our best fit to our measurements of
the $\mu_I$ dependence of $\beta_c$ at $m=0.035$ from our $dt=0.02$ runs gives
\begin{equation}
\beta_c \approx 5.15193 - 0.1758 \mu_I^2.
\end{equation}
This is in reasonable agreement with the estimate at finite $\mu$ of
de Forcrand and Philipsen \cite{deForcrand:2002ci,Philipsen}, 
if we make the suggested replacement $\mu_I = 2\mu$. 

\section{Discussion and Conclusions}

We are performing simulations of 3-flavour lattice QCD, using the hybrid
molecular-dynamics (R) algorithm, at small isospin chemical potential, close to
the finite temperature transition from hadronic/nuclear matter to a quark-gluon
plasma. The quark mass is chosen to be close to the critical value for zero
chemical potentials. The $\beta$ and hence temperature of this transition 
decrease slowly with increasing $\mu_I$, in a manner consistent with the 
decrease with increasing $\mu$, provided one identifies $\mu_I=2\mu$.

The nature of the transition is determined using Binder cumulants for the
chiral condensate. We find that the these cumulants depend strongly on the
updating increment $dt$, decreasing with increasing $dt$. This can be
understood from the fact that the shift in effective $\beta$ at finite $dt$ is
much larger below the transition than above it. Hence a small change in
$\beta$, which takes the system through the transition, produces a much larger
change in the effective $\beta$. This in turn induces a larger change in
observables, which makes the transition appear more abrupt, which is reflected
in a smaller value for $B_4$. When this dependence is taken into account, the
critical mass is found to be some 20\% below previously published values. This
is in agreement with the new results obtained by de Forcrand and Philipsen
using RHMC (exact) simulations \cite{Philipsen}.

When we apply the finite $dt$ corrections to our measurements of the Binder
cumulants at finite $\mu_I$, we find that they show little dependence on
$\mu_I$. So far, we find no evidence for the predicted critical endpoint. This
too is in agreement with the work of de Forcrand and Philipsen, whose new 
predictions of the nature of the transition at finite $\mu$ obtained from
continuations from imaginary $\mu$ also show a weak dependence of $B_4$ on
$\mu$ \cite{Philipsen}. 
In fact, their predictions suggest that $B_4$ might actually increase
with $\mu$ as was suggested by some of our early results.

We are continuing our simulations at a quark mass of $m=0.03$ which is closer
to the newly determined critical mass, and at $m=0.025$ which appears to be
just below the critical value. We too are converting to the new RHMC algorithm
\cite{Kennedy:1998cu},
which is exact. That is, it produces results which do not have any finite $dt$
errors.

If no critical endpoint is found, it will indicate that the critical endpoint,
if it exists, is unrelated to that found at zero chemical potentials, as the
mass is varied.

%The measurements that we are making in these simulations will also enable us
%to determine the equation-of-state for QCD at finite temperature and small
%$\mu_I$. It will be interesting to compare this with the corresponding form
%at finite $\mu$.

We intend to use simulations at finite $\mu_I$ as a platform for reweighting
(by the fermion phase factor) to finite $\mu$. This has been found to be a 
better choice than reweighting from zero $\mu$. Presumably this is because the 
position of the finite temperature phase transition is the same (or at least
close) for the 2 theories. In this respect it has similarities with the method
of reweighting from shifted $\beta$ values, used by Fodor and Katz. Here we
expect that new methods for simulating fermions will give us at least stochastic
estimators for (the phase of) the fermion determinant, removing the burden of
exact determinant calculations, which are too expensive.

\section*{Acknowledgements}
We thank P.~de~Forcrand and O.~Philipsen for helpful discussions. These
simulations are being run on the Jazz cluster at Argonne's LCRC, the Tungsten
and Cobalt clusters at NCSA, and the Jacquard cluster at NERSC. Some of the
small lattice runs were done on Linux PCs in the HEP division at Argonne.

\end{document}